\documentclass[fleqn,usenatbib]{mnras}

\usepackage{newtxtext,newtxmath}
\usepackage{subcaption}

\usepackage[T1]{fontenc}
\usepackage{xcolor} 

\DeclareRobustCommand{\VAN}[3]{#2}
\let\VANthebibliography\thebibliography
\def\thebibliography{\DeclareRobustCommand{\VAN}[3]{##3}\VANthebibliography}

\usepackage{graphicx}	
\usepackage{amsmath}	

\title{Vortex Dynamics in Various Solar Magnetic Field Configurations}

\author[A. Kannan and N. Yadav]{
Arjun Kannan$^{1}$ and Nitin Yadav$^{1}$\thanks{E-mail: nitnyadv@gmail.com}
\\
$^{1}$School of Physics, Indian Institute of Science Education and Research Thiruvananthapuram, Vithura, Thiruvananthapuram 695551, Kerala, India \\
}

\date{Accepted XXX. Received YYY; in original form ZZZ}

\pubyear{2024}

\begin{document}
\label{firstpage}
\pagerange{\pageref{firstpage}--\pageref{lastpage}}
\maketitle

\begin{abstract}
We investigate vortex dynamics in three magnetic regions, viz., Quiet Sun, Weak Plage, and Strong Plage, using realistic three-dimensional simulations from a comprehensive radiation-MHD code, MURaM. 
We find that the spatial extents and spatial distribution of vortices vary for different setups even though the photospheric turbulence responsible for generating vortices has similar profiles for all three regions.
We investigate kinetic and magnetic swirling strength and find them consistent with the Alfv\'en wave propagation.
Using a flux tube expansion model and linear magnetohydrodynamics (MHD) wave theory, we find that the deviation in kinetic swirling strength from the theoretically expected value is the highest for the Strong Plage, least for the Weak Plage, and intermediate for the Quiet Sun at chromospheric heights.
It suggests that Weak Plage is the most favoured region for chromospheric swirls, though they are of smaller spatial extents than in Quiet Sun.
We also conjecture that vortex interactions within a single flux tube in Strong Plage lead to an energy cascade from larger to smaller vortices that further result in much lower values of kinetic swirling strength than other regions. 
Fourier spectra of horizontal magnetic fields at 1 Mm height also show the steep cascade from large to smaller scales for Strong Plage. 
These findings indicate the potential of vortex-induced torsional Alfv\'en waves to travel higher in the atmosphere without damping for weaker magnetic regions such as the Quiet Sun, whereas vortices would result in dissipation and heating due to the vortex interactions in narrow flux tubes for the strongly magnetized regions such as Strong Plage. 
\end{abstract}




\section{Introduction}

Solar vortex flows are rotating plasma structures widely detected in direct solar observations as well as comprehensive numerical simulations (\cite{2023SSRv..219....1T}).
The detection of these swirling motions via high resolution observations trace back several decades (\cite{1988Natur.335..238B, 1995ApJ...447..419W}). 
In the recent years, a multitude of studies, both numerical and observational, have shown their significance in mass as well as energy transport and heating (\cite{2009A&A...507L...9W, 2011AnGeo..29..883S,2012A&A...541A..68M,2016A&A...586A..25P}).   
They are now abundantly detected in recent observations and numerical simulations, thanks to the latest advancements in instrumentation and high-performance computational facilities (\cite{2018ApJ...869..169G, 2019ApJ...881...83S, 2020ApJ...894L..17Y}). 
These vortex flows could be of either magnetic or non-magnetic nature in the near-surface layers (\cite{2012PhyS...86a8403K, 2016A&A...596A..43C}).
Nonetheless, in the higher solar atmospheric layers, the swirling structures are believed to be of magnetic nature (\cite{2012Natur.486..505W,2020A&A...639A..59M}). 
Turbulent convection beneath the solar surface excites perturbations at a large range of spatial and temporal scales (\cite{1998ApJ...499..914S,2003matu.book.....B, 2011MNRAS.416..148V}).
Since vortices are a natural ingredient of fluid turbulence, vortex flows are a common occurrence in the turbulent plasma fluid in the intergranular lanes as well.
They have the potential to perturb the footpoints of magnetic fields rooted at the surface and excite MHD waves (\cite{2013SSRv..175....1M,2013ApJ...776L...4S,2023RvMPP...7...17M}).
In the presence of magnetic fields permeating through the solar atmosphere, these perturbations travel to larger heights in the form of MHD waves, particularly torsional Alfv\'en waves (\cite{2013ApJ...768...17M, 2023LRSP...20....1J}).

Substantial evidences exist in the literature indicating the role of the Alfv\'en waves in energy transport and heating of the solar atmosphere (\cite{1978SoPh...56..305H,1981SoPh...70...25H,1982SoPh...75...35H,2018ApJ...869..169G}).
Using the high-resolution observations obtained by the IMaX instrument onboard balloon-borne SUNRISE telescope (\cite{2011SoPh..268....1B,2011SoPh..268...57M}), \cite{Bonet_2010} visually detected several small-scale vortices of magnetic nature.
\cite{2010A&A...513L...6B} reported on small-scale magnetic concentrations being dragged to the centre of a prominent vortex structure identified as a moving bright point.
Using 3D radiative MHD simulations of magnetoconvection, \cite{2011A&A...526A...5S} compared the properties and source terms of magnetic and non-magnetic vortices. 
They showed that vorticity is generated more efficiently in the magnetised model with magnetic tension in the intergranular magnetic flux concentrations being the dominant source term.
\cite{2013ApJ...776L...4S} showed that swirling motions at the intergranular lanes could also excite torsional Alfv\'en waves.

Using observations from SDO/AIA (\cite{SDO}) and from SST/CRISP (\cite{2003SPIE.4853..341S, 10.1007/978-3-642-02859-5_11}), \cite{2012Natur.486..505W} identified swirls at various heights in a coherent magnetic structure extending from solar surface to chromospheric heights.
They also performed numerical simulations demonstrating Alfv\'en waves being excited by photospheric vortex flow and directing the associated Poynting flux in the vertical direction.
Using HINODE and SST observartions, \cite{2019NatCo..10.3504L} reported the presence of ubiquitous Alfv\'en pulses excited by photospheric intensity swirls and reaching chromospheric heights carrying energy flux sufficient to balance the radiative energy losses.
Using the CO5BOLD MHD code \cite{2021A&A...649A.121B} found a high correlation between vortical motions and torsional magnetic field perturbations resulting in unidirectional Alfv\'en pulse signatures. 
They also conjectured that the majority of Poynting flux associated with these Alfv\'en pulses is due to complex superposition of small-scale vortical motions. 
This is also shown to be the case in the lower coronal heights using realistic simulations of a coronal loop by \cite{2023A&A...675A..94B}. 
Alfv\'en waves are crucial not only in energy transport from the photosphere to the upper layers but also in the heating of the chromosphere.
\cite{2011ApJ...736....3V} proposed Alfv\'en wave turbulence caused by counter-propagating waves as a possible physical mechanism heating the chromospheric plasma.
\cite{2017ApJ...840...20S} did a numerical investigation employing a torsional Alfv\'en wave driver to a cylindrical thin magnetic flux tube model and suggested Ohmic dissipation as the dominant heating mechanism in the lower and middle chromosphere.

Vortices in general arise self-consistently out of a pure hydrodynamical treatment and it is not necessary for magnetic fields to be present. Naturally, the question of differences between non-magnetic and magnetic vortex structures tends to arise, especially in the context of the Solar atmosphere. 
A numerical study done by \cite{2012A&A...541A..68M} on the comparison of magnetic and non-magnetic vortices revealed that non-magnetic vortices tend to primarily have a much lesser vertical extent, whereas the vortices coupled with the magnetic flux concentrations at the surface tend to reach chromospheric heights. 
They also report that both the cases contribute to local heating, but the dominating mechanism for the non-magnetic case is due to shocks, and for the magnetic case it is due to ohmic dissipation. 
There are also cases where non-magnetic and magnetic vortices are reported to be formed independently from each other. 
\cite{2021ApJ...915...24S} used the instantaneous vorticity deviation (IVD) and integrated averaged current deviation (IACD) methods to detect kinetic and magnetic vortices (dubbed as K and M vortices), respectively and showed that the K vortices tend to form at regions where the plasma-$\beta$ < 1 and magnetic vortices are formed at regions where plasma-$\beta$ > 1. 
These above-mentioned studies suggest that coupling of rotational motions with magnetic fields are favoured in certain magnetic conditions and these coupled kinetic and magnetic vortices are crucial for energy transport and heating of upper solar atmospheric layers.

Vortices are observed and detected in simulations in a wide range of scales, both spatially and temporally in the Solar atmosphere.
Observations typically report on comparatively large-scale vortices due to limitations in spatial resolution, whereas, simulations report vortices on much smaller scales as most vortex detection methods are based on velocity gradients that are stronger on smaller-scales.
Using phase difference analysis, \cite{2020A&A...643A.166T} observed a vortex flow driven primarily by the fast kink wave mode, but additionally also suggest the existence of localised small-scale Torsional Alfv\'en wave modes within the larger vortex.  
\cite{2020ApJ...894L..17Y} used the MURaM code for high-resolution simulations of a Plage region and reported on the multi-scale nature of vortices.
They demonstrated larger clusters consist of clusters of smaller vortices and showed that smaller vortices contribute more to energy transport in comparison to the larger vortices.
Quantitatively, vertical Poynting flux transported via small-scale vortex motions is shown to be adequate to compensate for the radiative losses in the chromosphere. 
Further \cite{2021A&A...645A...3Y} showed that there is significant enhancement in heating (viscous+resistive) over vortices in comparison to non-vortex regions, thereby suggesting their involvement in chromospheric heating.

In view of the reported association between kinetic and magnetic vortices, it is crucial to investigate how the background magnetic field configuration contributes to enabling the coupling between these two types of vortices.
Therefore, in this study, we conduct a comparative analysis of three distinct magnetic configurations, namely Quiet Sun, Weak Plage, and Strong Plage, with the objective of identifying Alfv\'enic signatures in the vortices.
After this introduction, we discuss the simulation setup and methods in Section \ref{sec:Methods}, followed by results and discussions in Section \ref{res_disc}, and then we summarize our key findings and conclude in Section \ref{conc}.

\begin{figure*}
	\includegraphics[width=2\columnwidth]{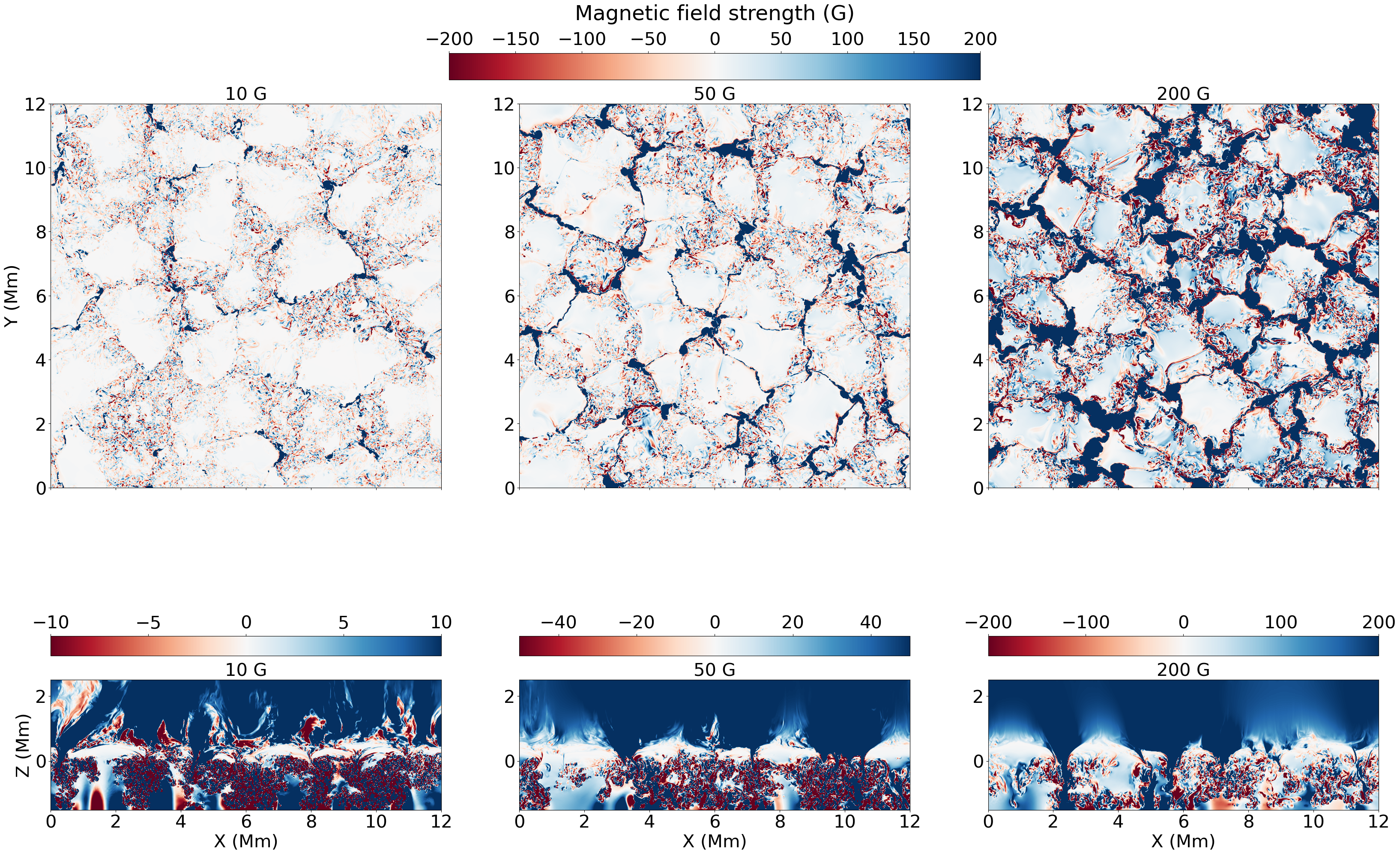}
    \caption{Maps showing slices of vertical magnetic field ($B_{z}$) for the three configurations. Top three images are in the plane z = 0 Mm, and bottom three are in the plane y = 10 Mm }
    \label{fig:bz_all}
\end{figure*}

\section{Methods}\label{sec:Methods}

\subsection{Simulation setup}
To simulate the various solar regions viz., Quiet Sun,
Weak Plage and Strong Plage, a three-dimensional radiative MHD code, MURaM is used.  
The MURaM code solves the MHD equations for a compressible, partially ionized plasma in a Cartesian box and employs a box-in-a-star approach to simulate the dynamics and energy transfer within the upper convective zone and the lower solar atmosphere. 
For a more comprehensive understanding of the equations governing the simulation and their numerical implementation, readers are referred to \cite{2005A&A...429..335V} and  \cite{2014ApJ...789..132R,2017ApJ...834...10R}.
A hydrodynamic (HD) run for two hours was followed by a MHD run for 1.2 hours until a quasi-static equilibrium state is achieved again. 
We tested this by analyzing the time evolution of the mean magnetic energy at each layer, ensuring its initial large scale oscillations have damped and oscillated only slightly around the mean position. 
This represents a physical state where the input energy is balanced by dissipation.
The MHD simulations were done for three different cases with mean vertical magnetic field strengths of 10 G, 50 G, and 200 G for simulating Quiet Sun, Weak Plage, and Strong Plage, respectively. 
Convection in the near surface layers is driven by the entropy fluctuations. Since the deeper layers have little influence on the small-scale granular flows at the surface, we can obtain physically consistent models using simulation boxes extending only a small fraction ($\sim$ 1 $\%$) of the whole convection zone by employing an open lower boundary condition.
The domain of the simulation box is 12 Mm $\times$ 12 Mm $\times$ 4 Mm (x, y, z, where z is normal to the solar surface, pointing outward) with a 10 km grid spacing in all the three directions. 
The lower boundary of the simulation box is situated 1.5 Mm below the mean solar surface.
Thus, our simulations incorporate realistic convective driving at the subsurface layers by turbulent convection, which is crucial for generating vorticity.
The top boundary is placed 2.5 Mm above the mean surface layer. It partially reflects the rotational motions, or torsional Alfv\'en waves, that reach it. 
This reflection impacts the dynamics in the layers near the top boundary. Therefore, we will focus our analysis on the layers only up to 1.5 Mm above the surface. 
Nonetheless, the reflection of Alfv\'en waves is not unreasonable, as the other footpoints of these magnetic fields would be anchored back in the photosphere in a real scenario, potentially exciting Alfvén waves that could enter our domain.
The boundary conditions used in these simulations are the same as in \cite{2020ApJ...894L..17Y,2021A&A...645A...3Y}.

\subsection{Vortex detection}
We use the Swirling Strength criterion as described by \cite{1990PhFlA...2..765C} and \cite{1999JFM...387..353Z} to detect regions exhibiting swirling motions in our data.
This detection method is advantageous over enhanced vorticity method as it only detects the pure rotation part and shear flows are not detected. 
This method involves eigen-analysis of the field gradient tensor, e.g., for velocity field $\mathbf{v}$, $A_{ij}=\partial_{i} v_{j}$ will take the following form:

\begin{equation}\label{eqn_grad}
\textbf{A} = 
\begin{pmatrix}
    \partial _{x} v_{x} & \partial _{x} v_{y} & \partial _{x} v_{z}\\
    \partial _{y} v_{x} & \partial _{y} v_{y} & \partial _{y} v_{z}\\
    \partial _{z} v_{x} & \partial _{z} v_{y} & \partial _{z} v_{z}\\
\end{pmatrix}
\end{equation}

In a region of swirling motion, $\textbf{A}_{ij}$ will have a conjugated pair of complex eigen-values. The matrix can then be diagonalised as follows:

\begin{equation}\label{eqn_matrix}
\textbf{A$^\prime$} = 
\begin{pmatrix}
    \lambda_{r} & 0 & 0\\
    0 & \lambda_{+} & 0\\
    0 & 0 & \lambda_{-}\\
\end{pmatrix}
\end{equation}

(where $\lambda_{\pm}$ = $\lambda_{cr} \pm i\lambda_{ci}$)

$\lambda_{ci}$ denotes the rotation rate of the local swirling motion of the field, and conventionally, the swirling strength is defined as $\lambda$ = 2$|\lambda_{ci}|$, with the associated time period of rotation being, $T = 4\pi / \lambda$.

\subsection{Convolution and threshold}\label{conv_thres}
\begin{figure*}
  \begin{subfigure}{0.5\textwidth}
    \centering
    \includegraphics[width=0.95\linewidth]{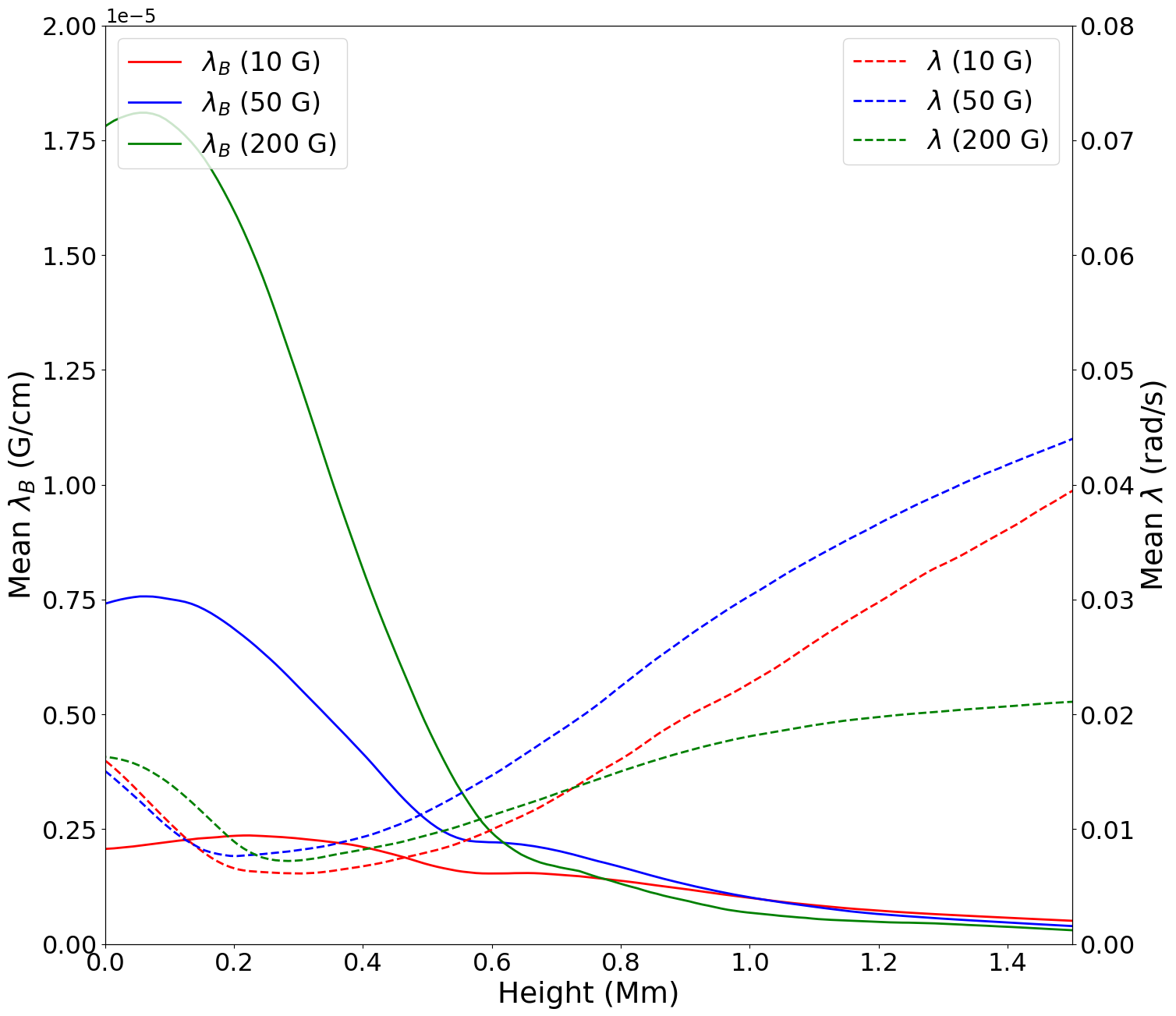}
    \caption{}
    \label{fig:ss_magss_h}
  \end{subfigure}%
  \begin{subfigure}{0.5\textwidth}
    \centering
    \includegraphics[width=0.95\linewidth]{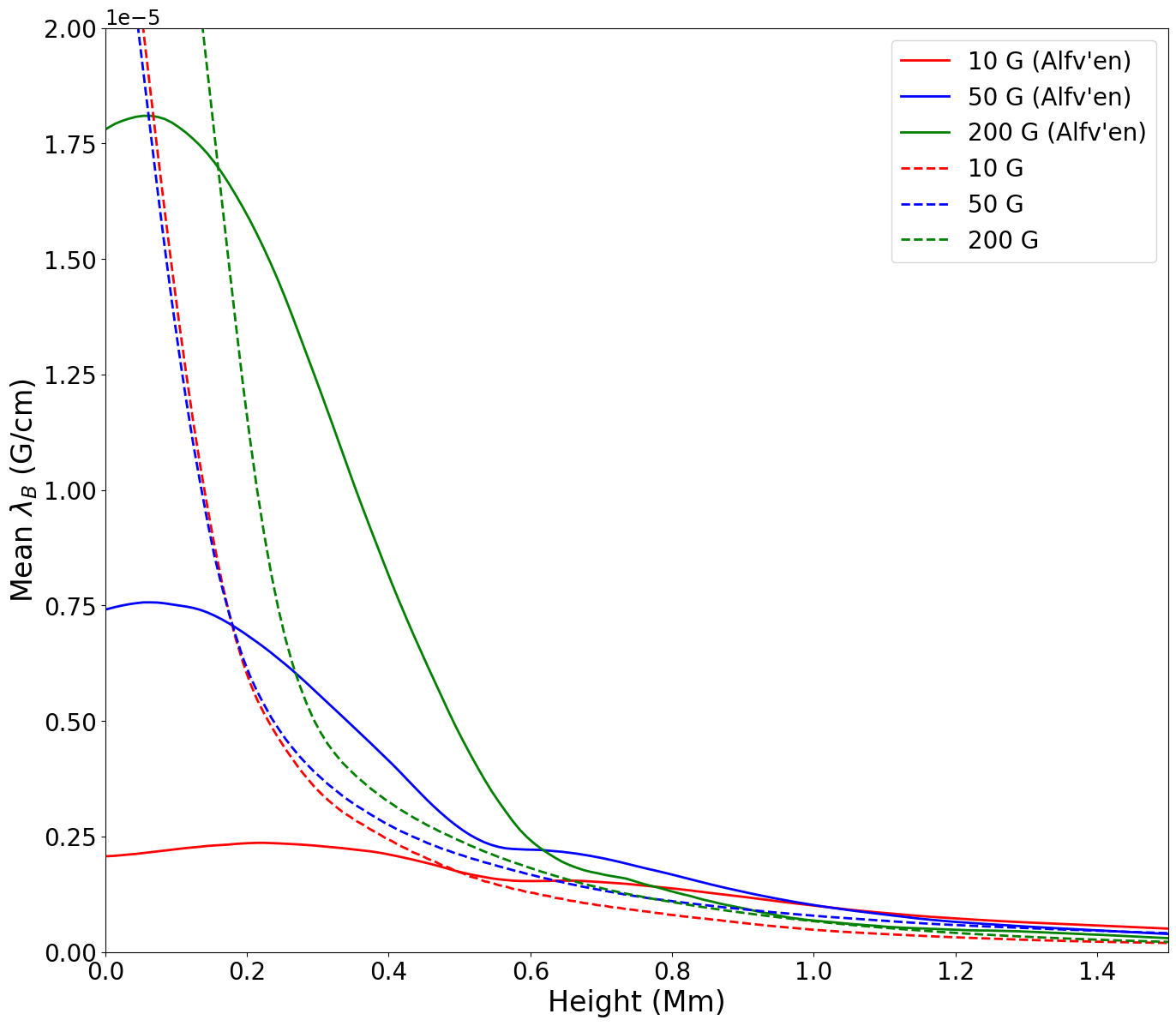}
    \caption{}
    \label{fig:magss_h}
  \end{subfigure}
    \caption{(a) Plot showing mean $\lambda$ and $\lambda_{B}$ over thresholded regions varying with height; the left vertical axis depicts values of $\lambda_{B}$ whereas the right one depicts values of $\lambda$. (b) Plot showing the comparison of $\lambda_{B}$ with the predicted value as a function of height}; the solid line is the same as in the left figure, whereas the dotted line is computed by the Alfv\'en wave relation over the thresholded Kinetic vortex regions.
  \label{fig:swstr_h}
\end{figure*}

\begin{figure*}
	\includegraphics[width=2\columnwidth]{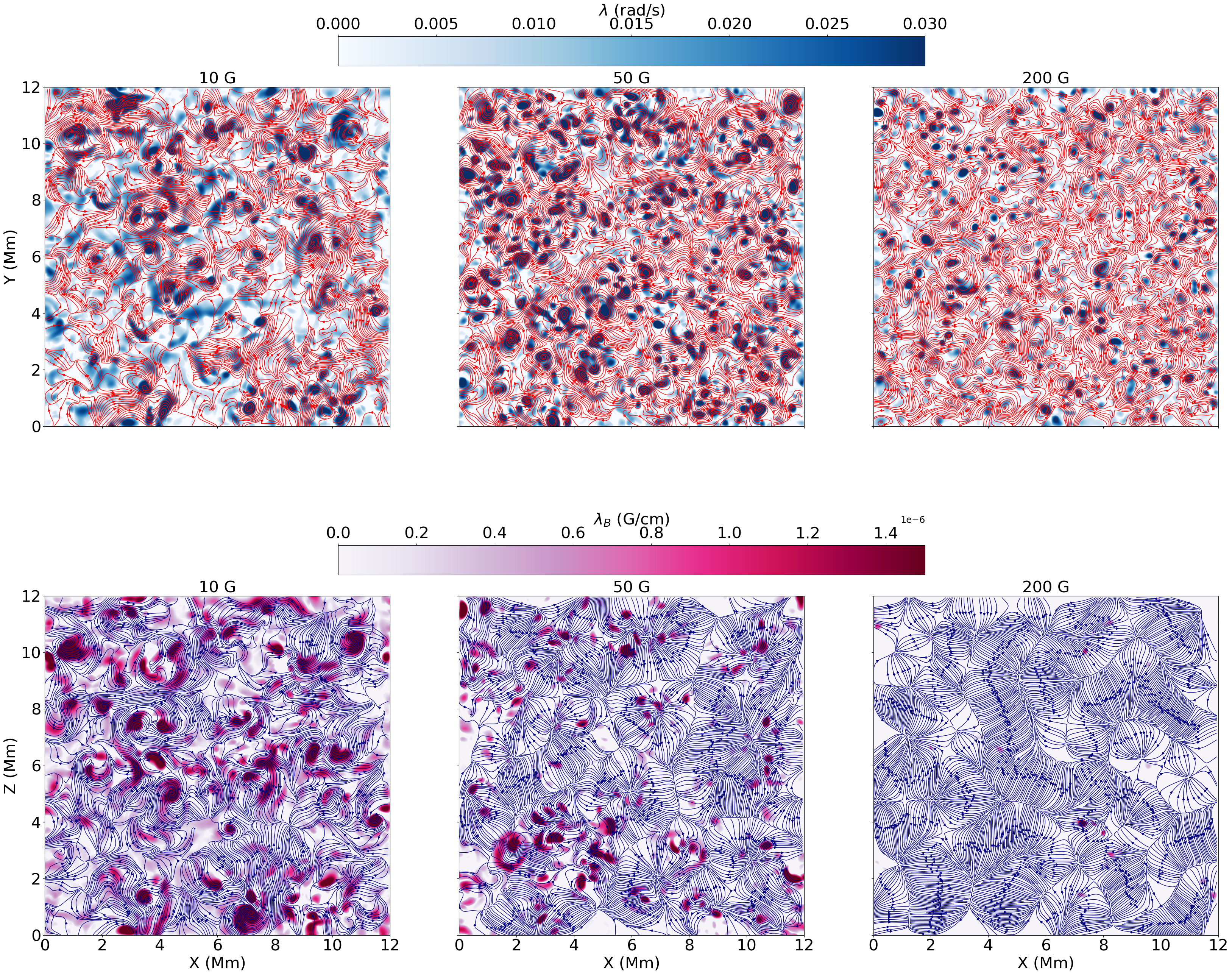}
    \caption{ Plots showing horizontal velocity (top) and magnetic field (bottom) streamlines projected at planes perpendicular to the vertical direction at a height, z = +1 Mm from the surface. The background maps are coloured by swirling strength and magnetic swirling strength respectively. }
    \label{fig:ss_all}
\end{figure*}

\begin{figure*}
  \begin{subfigure}{0.5\textwidth}
    \centering
    \includegraphics[width=0.95\linewidth]{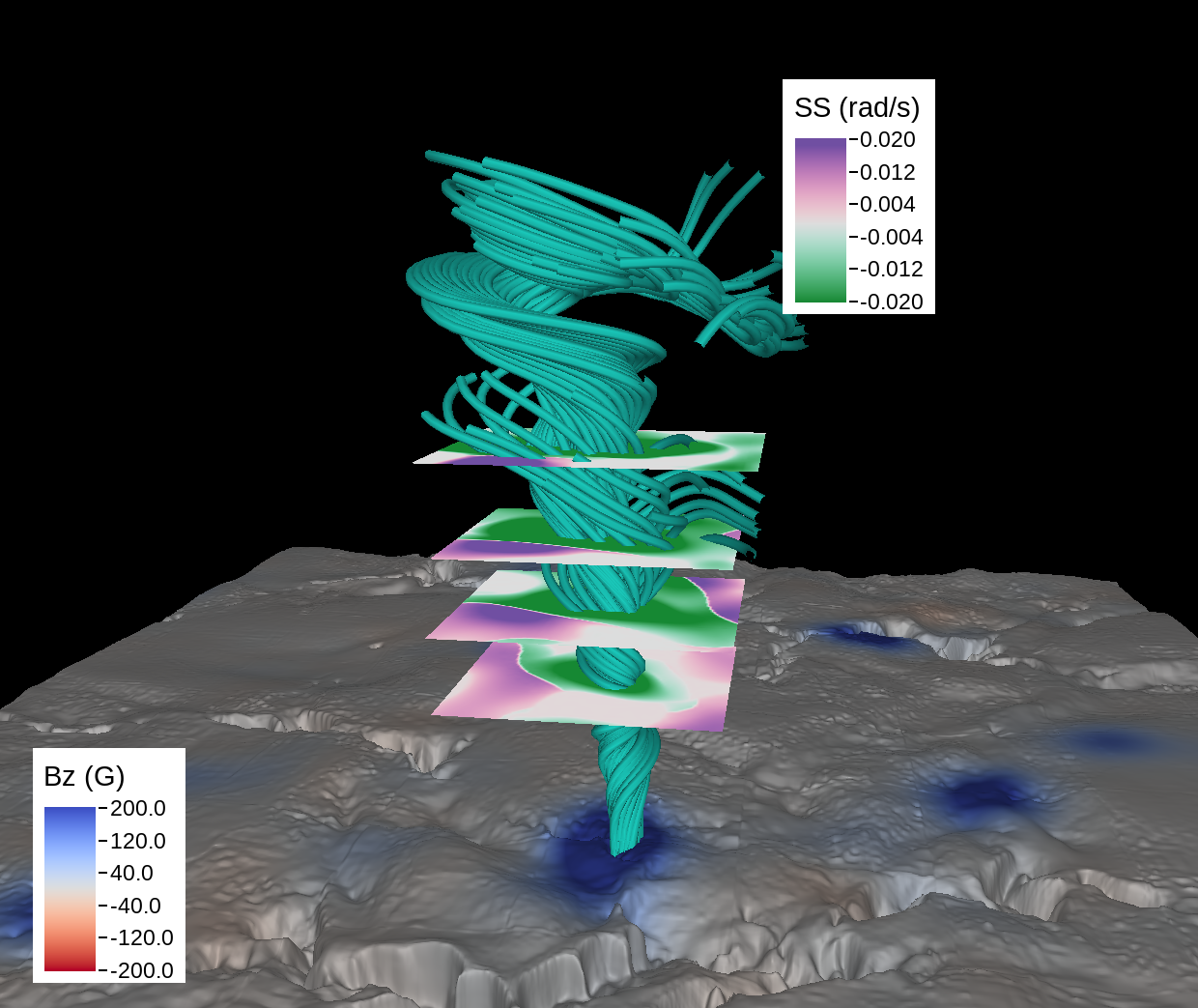}
    \caption{}
    \label{fig:vel_vort}
  \end{subfigure}%
  \begin{subfigure}{0.5\textwidth}
    \centering
    \includegraphics[width=0.95\linewidth]{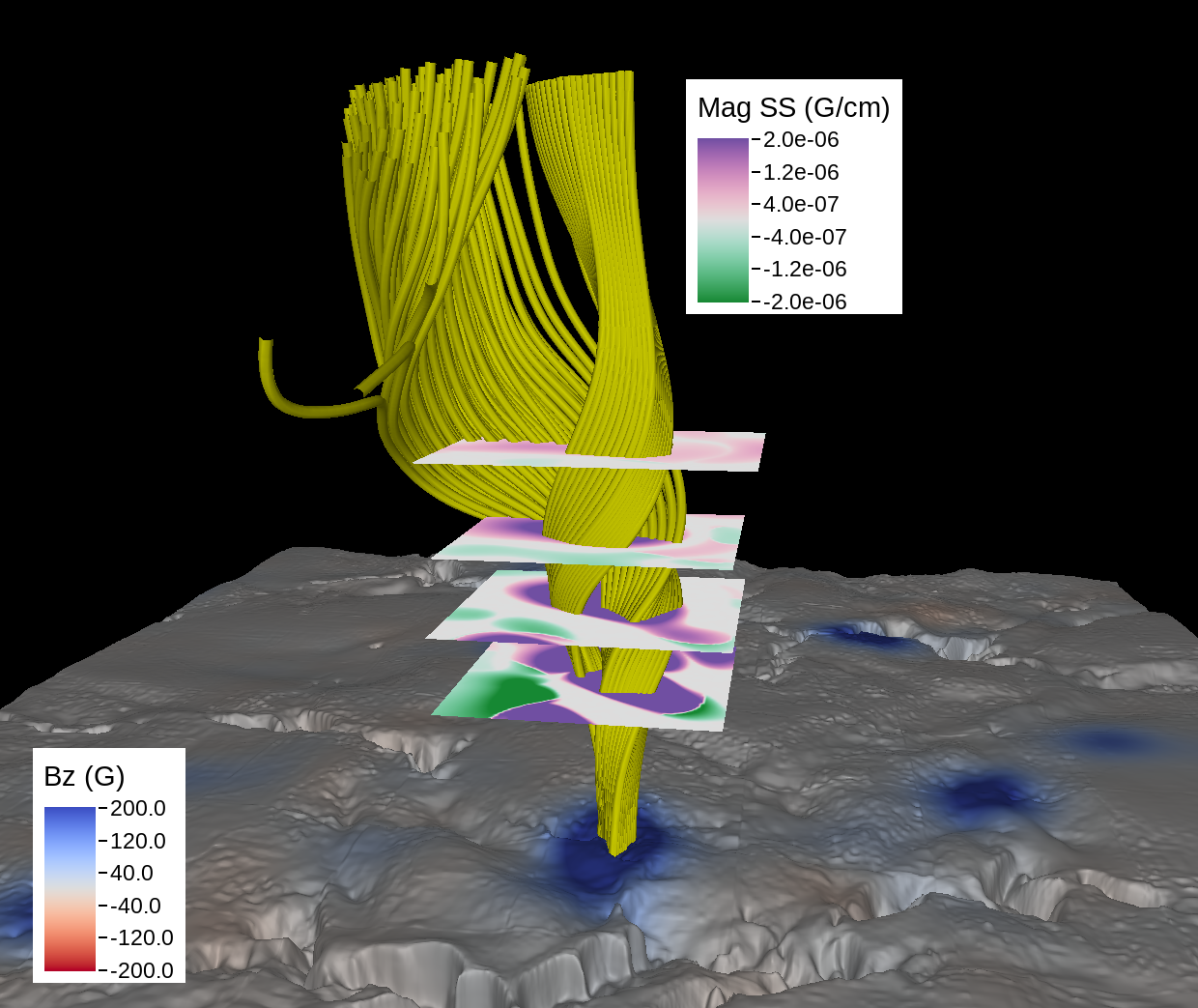}
    \caption{ }
    \label{fig:mag_vort}
  \end{subfigure}
  \caption{These images depict 3D streamplots showing a prominent vortex structure for the Quiet Sun case. In the left image, green lines depict velocity streamlines, and the horizontal maps show signed kinetic swirling strength at different heights. In the right image, yellow lines depict magnetic field streamlines, and the horizontal maps show signed magnetic swirling strength at various heights (purple and green colours represent counter-clockwise and clockwise rotation respectively). The gray surface is the $\tau$ = 1 layer, with the blue colours representing vertical magnetic field. }
  \label{fig:both_vorts}
\end{figure*}

Recently, \cite{2020A&A...633L...6K} reported the first direct observations of tosional Afv\'en waves in the solar corona using data from IRIS and SDO.
They exemplified on the possibility of excitation of torsional Afv\'en waves from magnetoconvective vortex flows that can twist and braid the magnetic fields.
They also suggested that such wave generation mechanism is likely to be very common as the lower solar atmosphere is filled with small-scale twisted magnetic structures.
Since the signatures of torsional Alfv\'en waves are more clear on larger scales, we degrade the spatial resolution of our high-resolution simulations and focus on comparatively large scale vortex flow and associated magnetic twist.
For this, all physical parameters are convolved with a Gaussian function having Full width at Half Maximum (FWHM) of 300 km. 
Our choice of FWHM is motivated by the available spatial resolution through current observations. Selecting a smaller (larger) FWHM would lead to smaller (larger) vortices. However, the overall conclusion drawn from the work remains unaffected.
The Swirling Strength is then computed on the convolved velocity and magnetic field vectors. 
The quantities depicted in Figures \ref{fig:swstr_h}- \ref{fig:flux_tubes} are obtained from the convolved data, while the remaining are derived from the full resolution data. 

Moreover, a threshold is applied for both $\lambda$ (calculated using velocity field) and $\lambda_{B}$ (calculated using magnetic field) that varies with height, and only those regions are considered for analysis that have swirling strengths above the threshold values. 
The threshold criteria is described as follows:  
$\lambda >$ $\mu + 0.5\sigma$ at each height; here $\mu$ and $\sigma$ are the mean and standard deviation of the distributions of log-scaled Swirling Strength at each height.
The threshold for swirling strength is determined through a trial-and-error process to ensure all visible rotational structures are accurately captured. Selecting a higher threshold only identifies the cores of rotational structures, failing to account for the complete swirling motion and resulting in a very small area fraction. Conversely, choosing a much lower threshold captures very weak rotational motions that may not be energetically significant.
From now on, whenever the term ``threshold" is referenced in the paper, it should be interpreted according to the preceding description.

\section{Results and Discussion} \label{res_disc}
\begin{figure*}
  \begin{subfigure}{0.5\textwidth}
    \centering
    \includegraphics[width=0.95\linewidth]{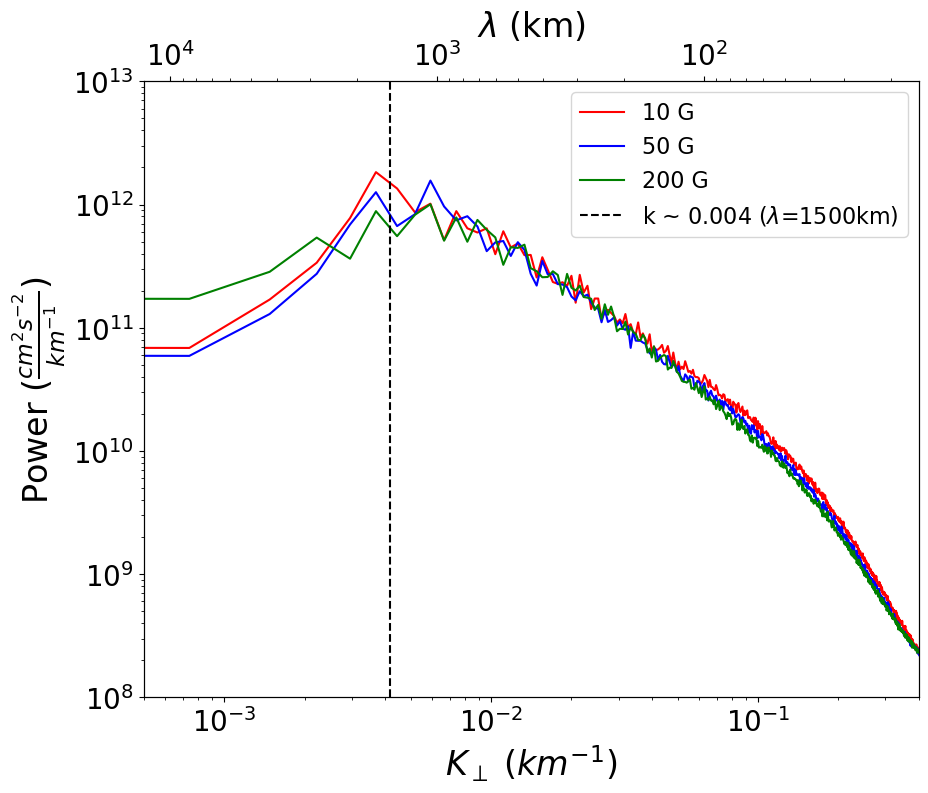}
    \caption{}
    \label{fig:vel_spectra}
  \end{subfigure}%
  \begin{subfigure}{0.5\textwidth}
    \centering
    \includegraphics[width=0.95\linewidth]{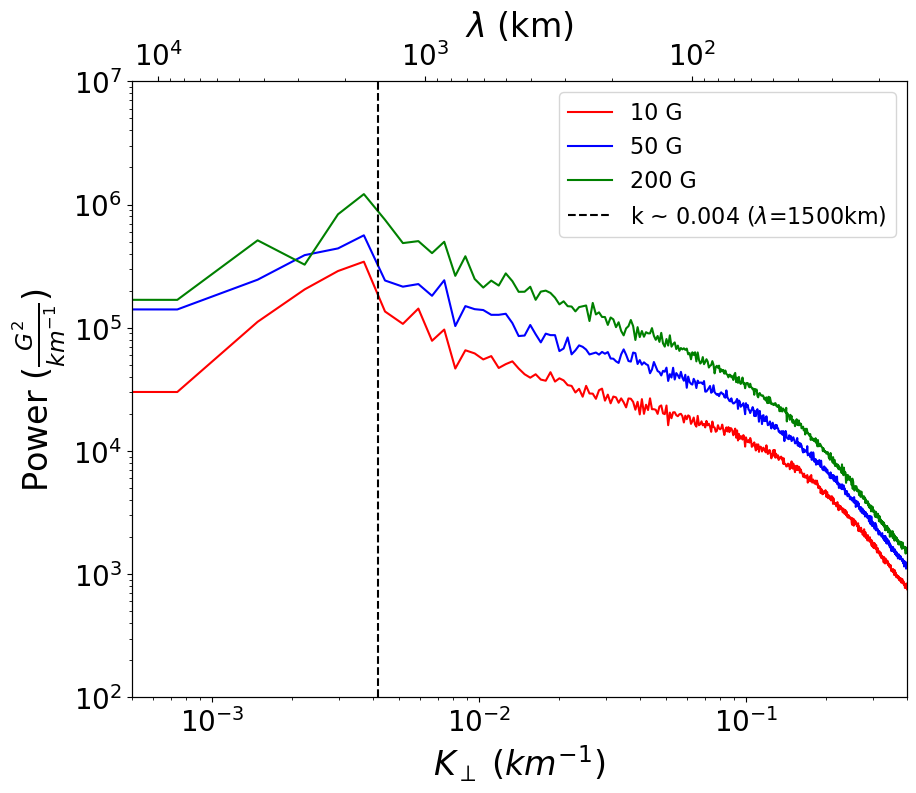}
    \caption{ }
    \label{fig:mag_spectra}
  \end{subfigure}
  \caption{Plots showing power spectra of horizontal velocity and horizontal magnetic field magnitude at the $\tau$ =1 layer. The horizontal axis represents wave number in the radial direction, and the dotted line corresponds to a radial length scale of about 1500 km.}
  \label{fig:spectras_tau1}
\end{figure*}
\subsection{Comparison of magnetic flux distribution}
We first compare the spatial distribution of the magnetic fields at the solar surface for the three sets of simulations that are representative of three magnetic regions, viz. Quiet Sun, Weak Plage, and Strong Plage. 
Figure \ref{fig:bz_all} shows the vertical component of the magnetic field strength vector at one instant for these three magnetic regions.
The top row corresponds to the mean solar surface, i.e., z = 0 Mm (xy-plane), and the bottom row corresponds to y = 10 Mm (xz-plane). 
It can be seen from the top row that the area coverage of the strong magnetic flux concentrations at the surface is highest for strong Plage followed by Weak Plage and Quiet Sun. 
There are also large areas covering small-scale mixed polarity fields in all the three setups. 
These magnetic fields form small-scale loops and do not extend to the higher atmospheric layers.
The magnetic field strength of mixed polarity fields is, on average, higher in Plage regions than in Quiet Sun regions.
From this, we gain insights into the magnetic field distribution on the solar surface, particularly at the intergranular lanes where vortices are usually excited due to turbulence.
The maps shown in the bottom row display the expansion of the magnetic flux tubes rooted in the intergranular lanes. 
It can be seen that the opposite polarity fields are more prominent at the chromospheric heights for the Quiet Sun case, whereas, in the weak and strong Plage cases, the magnetic field is mostly unipolar in the chromosphere. 
Hence, the magnetic field merging height also lies at a greater height from the surface in the Quiet Sun than in Weak Plage and Strong Plage. 
This magnetic field merging height is crucial in facilitating the interaction among flux tubes associated with different flux concentrations and generating in-situ vortices through chromospheric turbulence.
Magnetic fields are mostly turbulent in the layers below the mean solar surface in all three cases. 
Thus, the variation in the magnetic flux distribution on the solar surface and magnetic flux tube expansion affects the propagation and interaction of vortices, which will be explored and discussed in more detail later in Section \ref{flux_exp}. 

Next, we compared the distribution of the horizontal components of velocity and magnetic field at the corrugated  $\tau = 1$ layer (of the continuum at 500 nm), as they will be associated with vortex flows generated in the intergranular lanes having plasma downflows.
This plot indicates the variation in the driving of vortices at the solar surface for the three setups.
The left and right panels of Figure \ref{fig:spectras_tau1} show the power spectra of the horizontal velocity, $|v_{h}|$ and of the horizontal magnetic field, $|b_{h}|$ at the $\tau = 1$ layer, respectively. 
These are obtained after computing the discrete Fourier transform of the horizontal component of velocity and magnetic field data consecutively with respect to the X-axis and Y-axis.
Thereafter, the power in the radial direction was obtained by taking the magnitude of the spectra for each wavenumber bin in the form of thin rings, as described by \cite{2020A&A...644A..44Y}.
The velocity profile shows similar power for all three configurations as the plasma $\beta$ is more than unity at this layer, and therefore, the magnetic field does not play a significant role in the overall dynamics. 
However, as expected, the magnetic field spectra profile shows the highest power for the Strong Plage configuration, followed by Weak Plage and Quiet Sun regions. 
Therefore, even though kinetic vortices are formed similarly for all three configurations, the propagation of vortices to the higher layers may vary according to the magnetic field configurations, as vortex motions and magnetic field perturbations get coupled and propagate as torsional Alfv\'en waves.
To investigate it further, a comparison of Alfvén wave excitation in the three setups is done after the data is smoothed using a Gaussian filter in order to focus on large-scale swirling structures (as described in Section \ref{conv_thres}). 

\subsection{Comparison of $\lambda$ and $\lambda_{B}$ with height}
\begin{figure*}
   \begin{subfigure}{0.45\textwidth}
    \centering
    \includegraphics[width=0.95\linewidth]{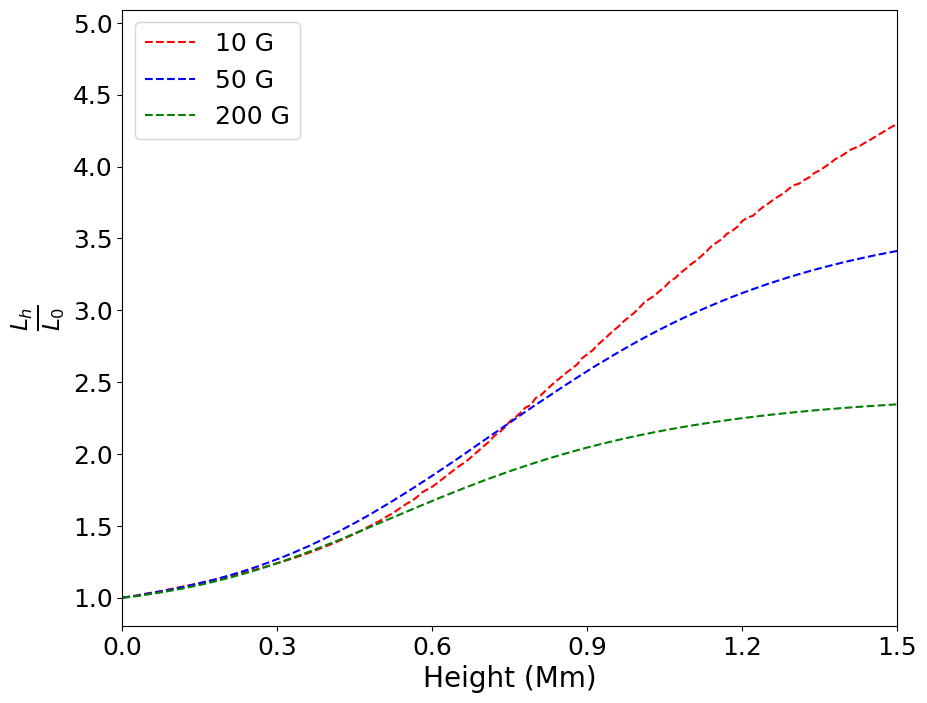}
    \caption{}
    \label{fig:tube_length_model}
   \end{subfigure}
  \begin{subfigure}{0.45\textwidth}
    \centering
    \includegraphics[width=0.95\linewidth]{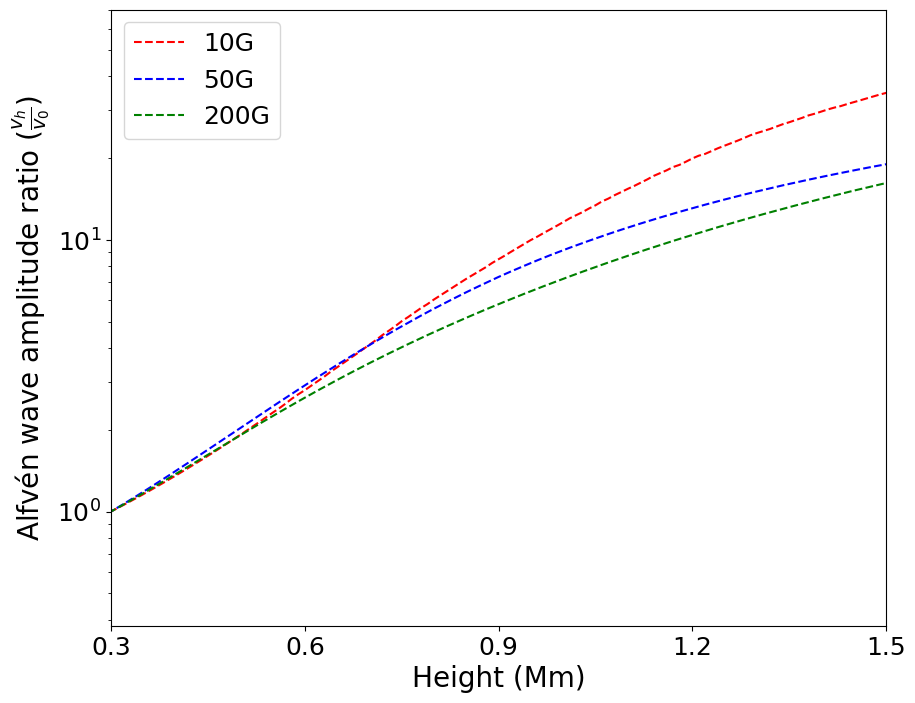}
    \caption{}
    \label{fig:v_osc_model}
  \end{subfigure}
  \begin{subfigure}{0.5\textwidth}
    \centering
    \includegraphics[width=0.95\linewidth]{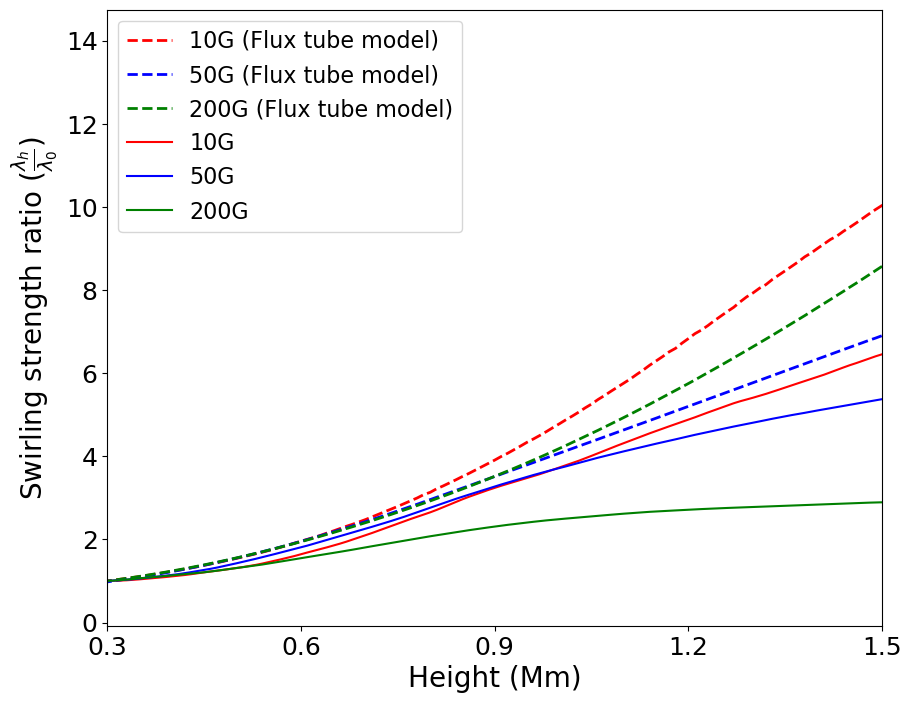}
    \caption{ }
    \label{fig:ss_model}
  \end{subfigure}
  \caption{Comparison with theoretically estimated values: (a) a plot of the Magnetic Flux tube length-scale at a height h to that at the surface height (z = 0 Mm) (b) a plot of Alfv\'en wave velocity amplitude ratio at height h to that at a height of z = 0.3 Mm from the surface. The calculation is done based on equation \ref{eqn2} for each height. (c) a plot showing the comparison of theoretically expected swirling strength (dashed lines) calculated as per equation \ref{eqn3}, and the actual values of swirling strength ($\lambda$) present in the simulation by taking a mean over the thresholded regions at each height (solid lines).}
  \label{fig:flux_tubes}
\end{figure*}

Figure \ref{fig:ss_magss_h} displays the variation of mean kinetic swirling strength $\lambda$ (dashed curves) and mean magnetic swirling strength $\lambda_B$ (solid curves) with height for all the three regions under investigation.
Here, the mean values are calculated over the locations satisfying the threshold criterion for vortex identification, as described in Section \ref{conv_thres}.
Different colors here correspond to different magnetic regions, i.e., Red, Blue, and Green colors for Quiet Sun, Weak Plage, and Strong Plage, respectively.
We see that the overall trend of kinetic swirling strength is similar for all three regions.
It tends to increase with height in general, apart from the dip close to the surface near the lower photospheric region. 
This increase in the swirling strength can be understood as an outcome of angular momentum conservation.
The mass density drops exponentially with height after the surface. 
Comparatively, the radii of vortices only increase on an almost linear scale following the magnetic flux tube expansion with height (as will be shown and discussed later in the description of Figure \ref{fig:tube_length_model}). 
Therefore, to compensate for the sharp density fall, rotational frequency increases.
The initial dip in the curve is associated with vortices existing only close to the surface, associated with the small-scale low-lying magnetic loops near the surface.
Vorticity propagates to higher layers via the virtue of magnetic field lines extending up to the top of the chromosphere.
On the other hand, vortices associated with small-scale fields do not propagate to higher layers, causing the dip in $\lambda$ right above the surface.
$\lambda$ starts increasing monotonically from about 0.2 - 0.3 Mm above the surface.  

As for $\lambda_B$, there is a general trend of higher magnitude near the surface region, followed by a continuous fall at larger heights. 
The initial increase could be due to the narrow magnetic flux tubes near the surface as they are squeezed by granular plasma flows, resulting in higher magnetic swirling strength. 
The fall in magnitude with height can be understood from the relation between the velocity and magnetic field perturbation for linear torsional Alfv\'en waves, given as follows: 
\begin{align}\label{alf_reln}
        |\textbf{b}| = |\textbf{v}|\sqrt{4\pi\rho}
\end{align}
Even though the magnetic field perturbations are proportional to velocity perturbations, the density of plasma drops exponentially with height due to the drop in atmospheric pressure. 
Therefore, though velocity perturbations only increase approximately on a linear/second-order scale with height, the magnitude of magnetic perturbations drops quickly with height.
Thus, the general trend of the obtained curves is consistent with angular momentum conservation and linear wave theory.
We performed further analyses to understand the variation among different magnetic regions, as described below.

To calculate the magnetic swirling strength consistent with linear wave theory over kinetic vortex regions, we compute the gradient of the RHS of equation \ref{alf_reln}.
Figure \ref{fig:magss_h} compares this theoretically estimated magnetic swirling strength over kinetic vortices against the magnetic swirling strength computed by Eigen-analysis as explained in Sec. \ref{conv_thres}.
In the layers near the solar surface, the value of $\lambda_{B}$ over the kinetic vortices, as expected from the linear wave theory, is much higher than the obtained $\lambda_{B}$ for all three configurations, possibly because a lot of kinetic vortices are generated in this region due to kinetic turbulence, which does not have a magnetic counterpart.
With increasing height, the two quantities are well matched for all three configurations, indicating that a significant amount of vorticity indeed propagates to the higher layers as Alfv\'en waves.   

Next, we compared the spatial distribution of kinetic and magnetic vortex regions for the three magnetic regions. 
Figure \ref{fig:ss_all} shows the comparison of velocity and magnetic field streamlines along with background maps of $\lambda$ and $\lambda_{B}$ at a plane perpendicular to the z-axis at 1 Mm height from the mean surface for all the three configurations. 
In the case of kinetic vortices, it can be seen that the vortex sizes are large for Quiet Sun (10 G) and much smaller for Strong Plage (200 G). For the Weak Plage case (50 G), they are of intermediate size, including both large-scale and small-scale vortices. 
The spatial extents of vortices seem to follow the same trend in the case of magnetic vortices. 
This region-dependent variation in spatial scales of vortices can be apprehended by the expansion of Magnetic flux tubes. 
The magnetic flux tube expansion will be highest in the Quiet Sun case and least in the Strong Plage case (as will be later shown and discussed in Fig \ref{fig:tube_length_model}) due to pressure balance. 
Therefore, vortex sizes at the chromospheric heights are expected to be smaller in stronger magnetic regions (Plage) than weaker magnetic regions (Quiet Sun), given the same scale size of vortex flows at the solar surface.

Another interesting observation is that the magnitude of swirling strength is lowest in the Strong Plage setup.
Whereas, if the rotational flow is not expanding much, then the swirling strength or the rotational velocity should be greater according to angular momentum conservation, given the similar amplitude of the velocity driver at the surface for all three configurations. 
This seems true for the Weak Plage compared to Quiet Sun but does not hold true for the Strong Plage, as their kinetic swirling strength is the lowest.
We conjecture that it is due to interactions among a large number of vortices originating in the same magnetic concentration.
This will result in a cascade of large to smaller vortices in strongly magnetic setups (as suggested by \cite{2021A&A...645A...3Y, 2020ApJ...894L..17Y}). 
That is possibly the reason for lower values of mean velocity swirling strength in the 200 G case, as their energy has been transferred to small-scale vortices due to interaction.
Since vortex-vortex interactions in a given magnetic flux tube will be the most dominant in the Strong Plage case among all magnetic regions, magnetic vortices will also be of much smaller scales and remain undetected. 
In the 50 G case, magnetic swirling strength is still detected at some places as vortex interactions are comparatively lesser than in 200 G, but at most places, it has disappeared due to interaction.

In the case of Quiet Sun (10 G), the opposite sense of rotation between plasma motions and the magnetic field twist is clearly seen for co-spatial locations, indicating the excitation and propagation of torsional Alfv\'en waves. 
In the case of Weak Plage (50G), the co-spatial rotations in the magnetic field are less prominent. There is a stark absence of magnetic field twist in the case of Strong Plage (200G) at 1 Mm. 
We suggest that due to vortex interactions, large-scale vortices decay into smaller-scale vortices and eventually undergo viscous and current dissipation, thus heating the plasma.
Thus, we conjecture that though more swirling motions are present in stronger magnetic regions near the surface, their detection might be difficult in chromospheric heights due to enhanced interaction and turbulence. 
Moreover, the twist in the magnetic field at these heights might be on a smaller scale in general for Plage regions due to Magnetic rigidity, which further explains the low area coverage of regions with high $\lambda_B$.  
This is also possibly the reason that chromospheric swirls (and magnetic twists) are mostly detected in Quiet Sun and, in particular, in coronal holes in direct solar observations.

Figure \ref{fig:both_vorts} shows a typical vortex flow in Quiet Sun data, where 3D velocity and magnetic field streamlines are plotted at a chosen vortex location. 
The planes indicate maps of $\lambda$ and $\lambda_{B}$ using the same color schemes, and the contrasting colors indicate an opposite sense of rotation in the maps.
The opposite sense of rotation is also visually evident from the streamlines.
Moreover, as expected, the structures are also rooted in a magnetic flux concentration at the surface ($\tau$ = 1 layer), again indicating torsional Alfv\'en wave excitation.

\subsection{Flux tube expansion and associated vortex flows}\label{flux_exp}
At the layers close to the solar surface, magnetic fields are swept away by the plasma flows in the intergranular lanes, resulting in magnetic flux concentrations. 
Here, the plasma downflows at the intergranular lanes have rotary motion and cause the winding up of the magnetic fields. 
Thus, vortical plasma motions can cause a twist in the flux tubes rooted into the intergranular lanes, exciting torsional Alfv\'en waves. 
An example of a vortex structure rooted at a magnetic flux concentration for the Quiet Sun configuration was shown in Figure \ref{fig:both_vorts}. 
A more generalised study analysing all magnetic flux concentrations and associated expanding magnetic flux tube with height is done as follows:
\begin{itemize}
    \item A suitable threshold for $B_{z}$ is chosen at the z=0 Mm (mean solar surface) to select strong magnetic concentrations such that their spatial scales are similar for all three configurations. This results in about 15-20 magnetic concentrations in each of the three cases. 
    \item For each flux concentration, points on the circumference of a circle centred at the maximum field strength location with a 100 km radius were chosen as seed points, and from there, magnetic field vector streamlines are calculated to extract associated magnetic flux tubes. 
    \item With the expansion of the Magnetic flux tube, the locations of the intersection of flux tubes with each height are interpolated, and these are the points where all physical quantities will be computed for further analysis. 
    \item The length scales at each height can be calculated using flux conservation along a Magnetic flux tube, as follows:

    \begin{align}\label{eqn_flux}
        L_{h} = \sqrt{\frac{\Phi}{B_{h}}}
    \end{align}

    Here, $\Phi$ is calculated at the reference layer (z = 0, in this case) for each concentration by multiplying the mean magnetic field at the boundary and the area of the circular region. $B_h$ is the mean magnetic field at a height h. Figure \ref{fig:tube_length_model} displays the length scale variation with height for the three regions under study. 
    Length scales as well as other quantities shown in the Figure \ref{fig:flux_tubes} are averaged over all selected magnetic concentrations.

    \item We define Alfv\'en wave velocity ratio as the ratio of horizontal velocity amplitude at the height h to the amplitude at z=0. It is calculated under the assumption that the transverse motions along a flux tube correspond to the torsional Alfv\'en waves that travel along these flux tubes and maintain linear nature, hence Alfv\'en wave energy flux remains constant (Eqs. \ref{eqn1} and \ref{eqn2}) as waves travel from surface to higher layers.
    The corresponding plot is shown in Figure \ref{fig:v_osc_model}.
    \begin{align}\label{eqn1}
        \rho v^{2} v_{A} = Const
    \end{align}
(Here, $\rho$ is density, $v$ is the velocity amplitude of Alfv\'en wave oscillation, and $v_{A}$ is the local Alfv\'en speed.)
    \begin{align}\label{eqn2}
        \cfrac{v_{h}}{v_{0}} = \sqrt{\cfrac{v_{A_{0}}\cdot\rho_{0}}{v_{A_{h}}\cdot\rho_{h}}}
    \end{align}
(The subscript 'h' refers to the quantity at a height h from the surface, whereas the subscript '0' represents the reference height that is used in order to calculate the ratio of the quantities.)
    \item A similar calculation can be done for the Swirling Strength ratio, where the theoretical value of Swirling Strength is obtained by dividing the velocity by the associated length scales at the corresponding heights, as shown in Eq. \ref{eqn3}. Figure \ref{fig:ss_model} displays the swirling strength ratio for the three setups.
    \begin{align}\label{eqn3}
        \cfrac{\lambda_{h}}{\lambda_{0}} = \cfrac{L_{0}}{L_{h}}\sqrt{\cfrac{v_{A_{0}}\cdot\rho_{0}}{v_{A_{h}}\cdot\rho_{h}}}
    \end{align}
    
\end{itemize}
Note that in Fig \ref{fig:tube_length_model}, the surface is chosen as the reference height, whereas for Figures \ref{fig:v_osc_model} and \ref{fig:ss_model} a height of 0.3 Mm from the surface is chosen. This reference height is chosen in order to discard vortices associated with mixed polarity magnetic regions, as explained in the description of Figure \ref{fig:ss_magss_h}.
Theoretically, the amplitude of the angular velocity of the fluid should remain much lower than the local Alfv\'en speed in the simulation domain under study, thereby preserving the linear wave nature throughout the domain. 
An increase in angular velocity and consequently in angular frequency is, however, expected even for linear wave propagation through a gravitationally stratified medium, as the wave energy flux remains conserved. 
It is crucial to note that wave frequency, which stays constant during linear wave propagation in the absence of nonlinearities and turbulence, should not be confused with this increase in angular frequency.
In our case these waves are excited by a angular velocity pulse, which can theoretically be expressed as a combination of various linear waves.  
As they propagate through an inhomogeneous medium and interact with oppositely propagating waves, their velocity power is distributed across different frequencies.
As a result, due to turbulence, pulse frequency would also change as it propagates to higher layers in addition to angular frequency.

\begin{figure*}
    \includegraphics[width=1.8\columnwidth]{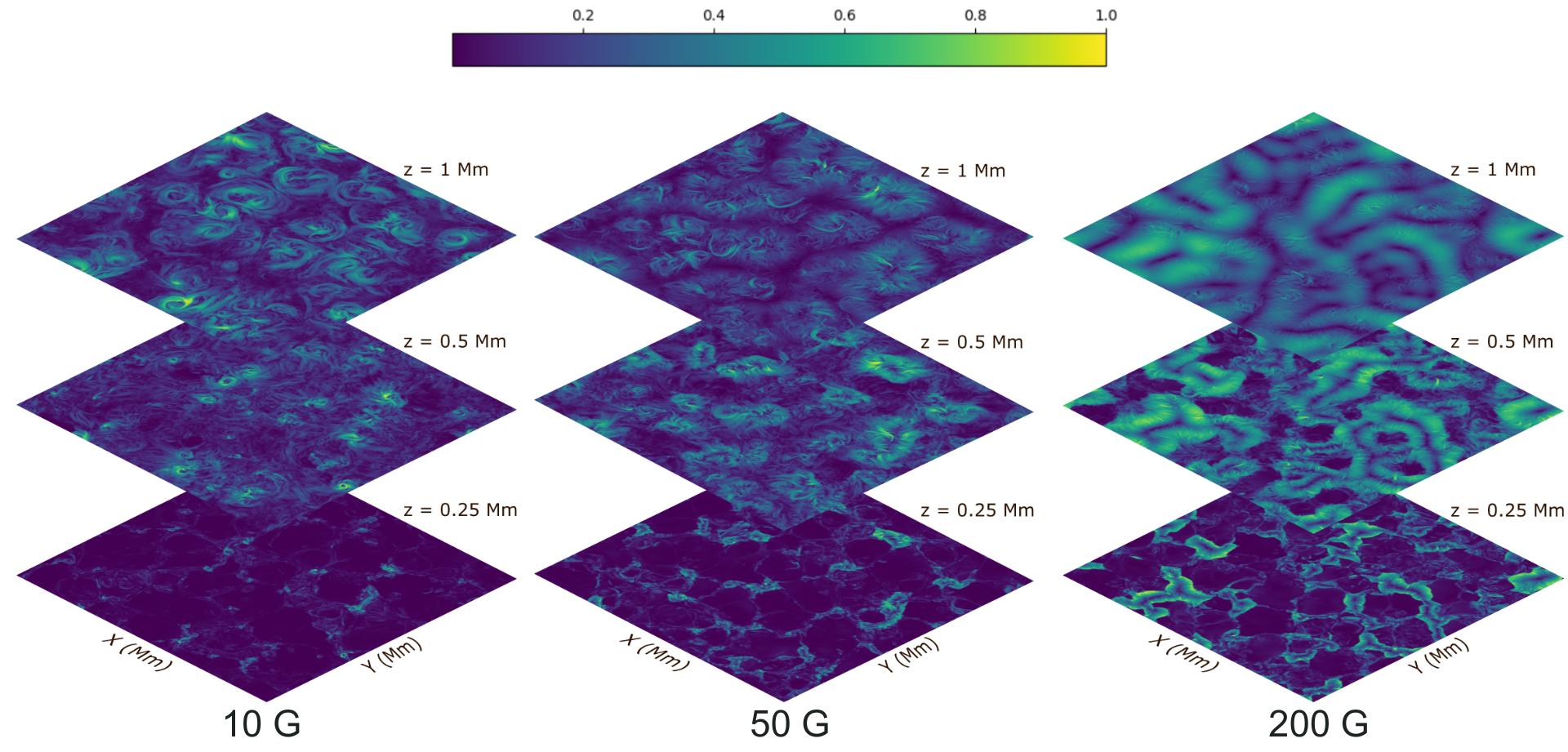}
    \caption{Maps (at planes perpendicular to the vertical direction) depicting magnitude of horizontal magnetic field for the three configurations at three different heights, as indicated in the side. The values are normalised by the maximum at each horizontal slice.}
    \label{fig:bh_all}
\end{figure*}

\begin{figure*}
	\includegraphics[width=2\columnwidth]{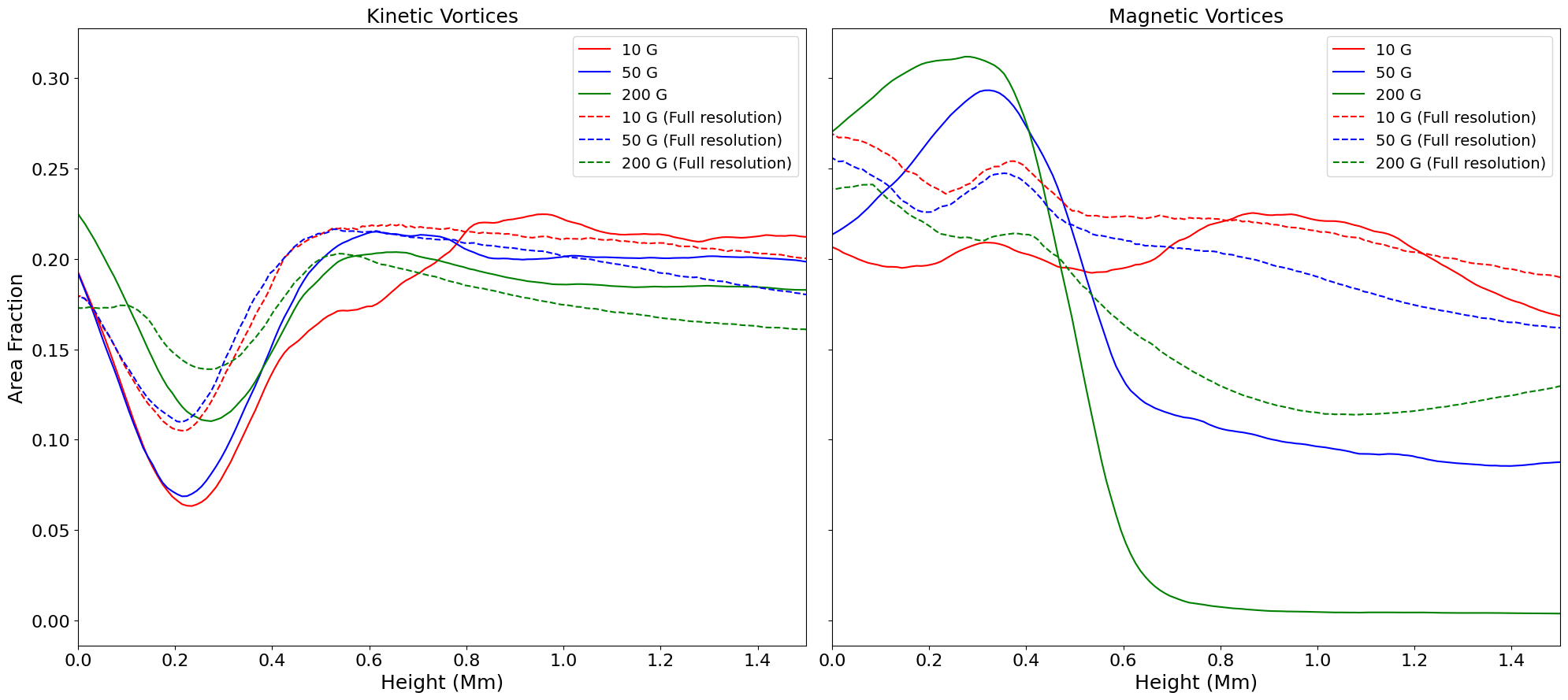}
    \caption{This figure shows the area fraction covered by kinetic and magnetic vortices satisfying the threshold criteria for both the convolved (solid lines) and full resolution data (dashed lines).}
    \label{fig:area_frac_fullres}
\end{figure*}

\begin{figure*}[h]
  \begin{subfigure}{0.5\textwidth}
    \centering
    \includegraphics[width=0.95\linewidth]{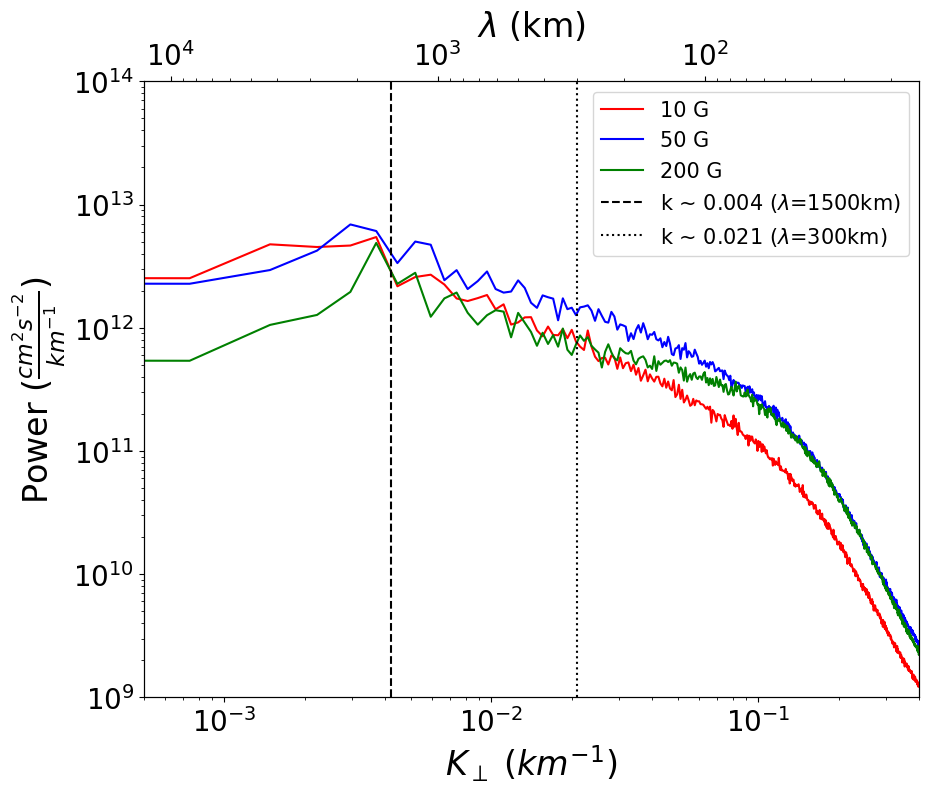}
    \caption{}
    \label{fig:vel_spectra1}
  \end{subfigure}%
  \begin{subfigure}{0.5\textwidth}
    \centering
    \includegraphics[width=0.95\linewidth]{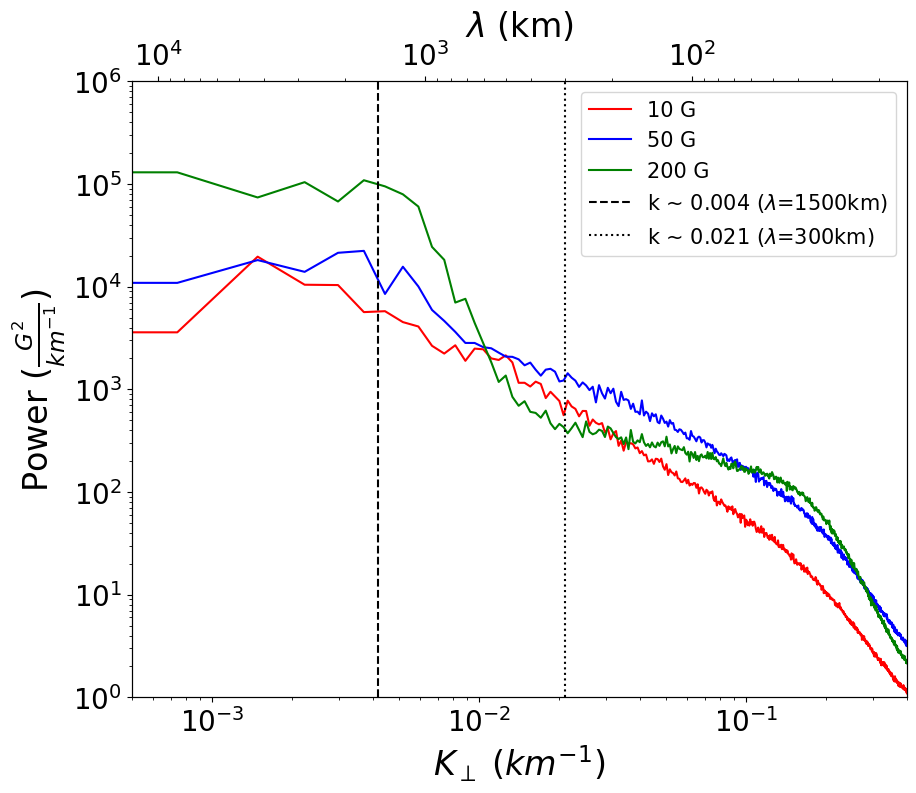}
    \caption{}
    \label{fig:mag_spectra2}
  \end{subfigure}
  \caption{Plots showing power spectra of horizontal velocity (left) and horizontal magnetic field magnitude (right) at the z = 1 Mm layer. The horizontal axis represents wave number in the radial direction, the dashed and dotted lines correspond to radial length scales of 1500 km and 300 km respectively. }
  \label{fig:spectras_z1}
\end{figure*}

The expansion in the magnetic flux tube is higher for the Quiet Sun case than the Plage regions as shown in \ref{fig:tube_length_model}. 
This can be understood from the thin flux tube model (\cite{1999A&A...347L..27S}).
We note that independent of the magnetic region under study, the magnetic flux concentrations at the surface have a very high magnetic field strength ($\sim$kG). 
With increasing height, thermal pressure drops due to gravitational stratification, and accordingly, magnetic pressure also drops. 
This drop in magnetic pressure leads to a decrease in the magnetic field strength, and flux tubes begin to expand out in order to maintain the constant flux and finally merge with other neighbouring flux tubes. 
The average height at which flux tubes merge is commonly referred to as merging height. After merging height, there isn't much further expansion in the flux tubes as the environment outside an individual flux tube is not field free anymore. 
Therefore, to compensate for the external total pressure (thermal + magnetic), the expansion has to be more in the case of Quiet Sun, followed by Weak Plage and then Strong Plage.
This expansion is also responsible for the vortices in general being larger in size for the Quiet Sun case as compared to the Plage cases as seen in Figure \ref{fig:swstr_h}. 

Moreover, from Figure \ref{fig:v_osc_model}, it is clear that if torsional Alfv\'en waves travel undisturbed in the atmosphere, its associated velocity perturbation amplitude increases with height for all three configurations, as estimated from Equation \ref{eqn2}. 
Next, Figure \ref{fig:ss_model} shows the theoretically estimated swirling strength ratio from flux tube model described above (dashed curves) along with the actual swirling strength ratio computed by eigen analysis as described in Section \ref{conv_thres} (solid curves).
It is clear that the actual swirling strength present in the chromosphere is lesser as compared to the theoretically calculated for all the three configurations, with the deviation being the highest for the Strong Plage case.

This detailed analysis based on theoretical model of flux tube expansion and propagation of linear torsional Alfv\'en waves supports our hypothesis that interaction between vortices within a single flux tube results in generation of small-scale vortices and energy cascade from larger to smaller vortices. 
It is worth pointing out here that for the Weak Plage case, the deviation is the least out of all the cases. 
This is possibly due to higher flux tube expansion in Weak Plage than Strong Plage, and larger areas of magnetic concentrations than the Quiet Sun, leading to the most favoured environment for Alfv\'en wave excitation and propagation in a flux tube with lesser vortex interactions. 
However, this cannot rule out interactions among vortices as there is a clear drop in the area covered by magnetic swirls for the Weak Plage case (as compared to Quiet Sun) as seen in Figure \ref{fig:ss_all}. 
From this analysis, it is clear that variation in expansion of the flux tubes play a crucial role in determining the vortex interactions and cascade to smaller scales which further governs the efficient dissipation (resistive and viscous) and heating of the solar plasma. 
Thus, we show that contribution of vortices in heating of the solar atmosphere will vary in different solar regions. 
We conjecture that strong magnetic regions like Plage, having larger areas of magnetic concentrations, offer more sites for excitation of torsional Alfv\'en waves in comparison to Quiet Sun. 
Moreover, since expansion of magnetic flux tubes is lesser in strongly magnetic regions, it will lead to enhanced vortex interactions in a given flux tube and hence enhanced dissipation and heating.

To support our conclusion regarding enhanced cascade in stronger magnetic regions, we did further analysis using full resolution data as explained in the following section.

\subsection{Comparison with full resolution}
Figure \ref{fig:bh_all} shows the structure of horizontal magnetic field for all three configurations as the flux tubes expands with height. 
The swirling structures or magnetic field twist are prominently seen for the Quiet Sun case at a height of 1 Mm. 
For the Weak Plage case, the structures have a more complicated pattern at heights of 0.5 Mm and at 1 Mm, owing to the interaction between the individual swirling structures. 
In the case of Strong Plage, at a height of 0.5 Mm, there are swirling structures interacting with each other. 
However, at a height of 1 Mm, the structure is more or less homogeneous, with sparse small scale swirls. 
This is possibly resulting from enhanced vortex interactions, increased generation of smaller vortices and efficient dissipation due to resistive and viscous dissipation working more effectively on smaller scales.
This is also evident by the values of $\lambda_B$ at these heights for the Strong Plage case as seen in Fig \ref{fig:ss_magss_h}. 

Next, we compare the area fraction covered by the swirls in convolved data and the full resolution data as shown in Figure \ref{fig:area_frac_fullres}. 
In the case of area coverage by kinetic vortices (left), there is more area coverage in full resolution data, especially in the near surface layers (up to 0.3 Mm) where small-scale vortices associated with mixed-polarity fields dominate as they will be greatly affected by the degradation procedure.
However, in the higher layers convolution does not affect the area coverage much as vortex scale sizes increase due to the magnetic flux tube expansion. 
There is still some differences due to the generation of small-scale vortices by vortex interactions that are wiped out in the degradation procedure.

In the case of magnetic vortices (right panel of Figure \ref{fig:area_frac_fullres}), the area coverage in the Weak Plage follows a very similar trend as the Quiet Sun in full resolution data (dashed curves), with a lower area coverage in general across all heights as expected due to lesser flux tube expansion. 
Whereas, for Strong Plage, the area coverage drops more than that of Weak Plage, supporting the enhanced interactions between vortices in a flux tube and cascade to smaller vortices that do not have a magnetic nature. 
Degradation does not affect the area coverage much for the Quiet Sun as the vortex interaction will be least there. 
However, it discernibly affects the area coverage for Plage regions as magnetic vortices are smaller in Plage regions and, hence, more susceptible to degradation, resulting in large deviations in area coverage.

Finally, to verify the non-magnetic nature of chromospheric vortices that are generated in-situ by vortex interactions and do not have photospheric origin, we calculate power spectra of horizontal velocity and horizontal magnetic field at z = 1 Mm using the full resolution data, as shown in Figure \ref{fig:spectras_z1}. 
From the horizontal velocity spectra, both the Quiet Sun and Weak Plage show higher power in radial scales larger than 1500 km compared to the Strong Plage case. 
That can be understood as energy from larger scales has been cascaded to smaller scales due to vortex interactions.
As for the magnetic field spectra, the Strong Plage region exhibits higher power for scales larger than 1500 km, representing the homogeneous structure seen in Figure \ref{fig:bh_all}. 
However, for scales between 1500 and 300 km, the power for Strong Plage drops steeply, supporting our conjecture of generation of non-magnetic small-scale vortices by vortex interaction.
We call them non-magnetic vortices as they are formed in the chromosphere itself unlike other chromospheric vortices that have photospheric origin and reach to chromospheric height by the virtue of magnetic field and Alfv\'en wave propagation.
\section{Conclusions} \label{conc}
Vortices or rotational plasma motions are natural ingredients of a turbulent flow, and the turbulent plasma fluid in the intergranular lanes at the solar surface provides the perfect conditions for vortex formation. 
Vortices perturb the footpoints of magnetic fields anchored at the solar surface and excite torsional Alfv\'en waves that could transport the energy to the higher layers and potentially heat the solar plasma. 
Thus, a comparative investigation of vortices and associated magnetic perturbations is essential to probe their role in the excitation of torsional Alfv\'en waves and solar plasma heating in different solar regions.
In this paper, realistic three-dimensional simulations from a comprehensive radiation-MHD code, MURaM simulation are performed and are analysed to compare vortical motions and the nature of Alfv\'en wave excitation in three different magnetic field configurations, viz., Quiet Sun, Weak
Plage, and Strong Plage. 
The data is first degraded with a Gaussian filter to gauge the signatures of Alfv\'en waves at larger scales effectively. 
The swirling strength criterion is used to detect the locations of vortex flows and associated twists in magnetic fields for both the full-resolution and degraded datasets. 
The power distribution of the velocity and magnetic field horizontal components at $\tau$= 1 layer was examined, and we found that the velocity power spectra for all three setups are statistically identical. 
Consequently, there are no differences in the driving or excitation of vortices in the three solar regions.
Magnetic power spectra, however, had variations in power levels over all spatial scales.
Alfv\'en waves are closely linked to vortex motions due to the nature of magnetized plasma fluid and turbulent magneto-convection at the photospheric layers.
Torsional Alfv\'en waves are excited at the surface by rotational motions and propagate along magnetic fields to the chromospheric heights, as shown in the present study for all three magnetic field setups. 
These results are consistent with the findings of \cite{2021A&A...649A.121B}, where they report the velocity and magnetic perturbations propagating in a uni-directional pulse-like manner.
We verified their findings and extended them to different magnetic regions.
Since the magnetic field strengths, spatial distributions, and configurations are different for different regions, it results in variations in the propagation of vorticity or swirling strength and associated magnetic swirling strength.
We find that spatial extents of kinetic vortices are largest in the chromosphere for the Quiet sun and smallest for the Strong Plage.
Moreover, the mean magnitude of swirling strength is maximum for the region with intermediate magnetic strength, i.e., Weak Plage, whereas one would expect Strong Plage to have higher magnitudes of swirling strength.

To understand this discrepancy, we constructed a theoretical model of torsional Alfv\'en wave propagation along a flux tube. 
We find that lesser expansion of magnetic flux tubes in the case of the Strong Plage region as compared to Quiet Sun leads to a higher possibility of vortex flows interacting amongst each other, generating turbulence and hence leading to energy transfer from larger to smaller scales, in agreement as proposed by \cite{2020ApJ...894L..17Y,2021A&A...645A...3Y}. 
Comparison between the convolved and full-resolution data verifies that the energy cascade is the highest for the Strong Plage case at the chromospheric heights. 
Our findings suggest that vortices might contribute more towards chromospheric heating via interaction between vortex motions and Alfv\'en waves in Strong Plage regions, whereas in weaker magnetic regions, such as Quiet Sun, they would lead to a more effective propagation of Alfv\'en waves to higher atmospheric heights without a significant amount of damping.

\section*{Acknowledgements}
N.Y. acknowledges the support from the DST INSPIRE Faculty Grant (IF21-PH~268) and the SERB MATRICS grant (MTR/2023/001332).
N.Y. gratefully acknowledges the computational resources provided by the COBRA supercomputer of the Max Planck Computing and Data Facility (MPCDF) in Garching, Germany.

\section*{Data Availability}
Data sets generated during the current study are available from the corresponding author on a reasonable request. 


\bibliographystyle{mnras}
\bibliography{example} 










\bsp	
\label{lastpage}
\end{document}